\documentclass[fleqn]{2020SCGE} %, fleqn
\usepackage{siunitx}
\usepackage{hyperref}
\usepackage{booktabs}
\usepackage{lineno}
\usepackage{color}
\usepackage{verbatim}
\usepackage{float}

\usepackage{graphicx}
\usepackage{amsmath}
\usepackage{amssymb}	% Extra maths symbols
\usepackage{diagbox}
\usepackage{threeparttable}

\usepackage{ulem}
\usepackage[numbers]{natbib}
\usepackage{longtable, threeparttablex, booktabs, url}
\usepackage{multirow} % 支持合并单元格
\usepackage{array} % 改进表格功能
\usepackage{makecell} % 支持表格中的自动换行
\usepackage{colortbl}
\usepackage{xcolor}
\usepackage{tabularx} % 自适应表格宽度

\usepackage{color}
\begin{document}
%\linenumbers 
\ensubject{subject}
%%%%%%%%%%%%%%%%%%%%%%%%%%%%%%%%%%%%%%%%%%%%%%%%%%%%%%%
%%% Authors do not modify the information below
%Letter to the Editor
\ArticleType{Article}
\SpecialTopic{SPECIAL TOPIC: }
\Year{2025}
\Month{xxx}
\Vol{xx}
\No{x}
\DOI{xx}
\ArtNo{000000}
\ReceiveDate{xx}
\AcceptDate{xx}
%%% Author information for page head. 
\AuthorMark{Lei Zhang}

%%% Authors for citation. 
\AuthorCitation{ Lei Zhang, et al.}
%\OnlineDate{January 1, 2016}
%%%%%%%%%%%%%%%%%%%%%%%%%%%%%%%%%%%%%%%%%%%%%%%%%%%%%%%
\title{Sensitive Constraints on Coherent Radio Emission from \\Five Isolated White Dwarfs}

%%% Corresponding author: 
%%%   \author[number]{Full name}{{email@xxx.com}}
%%% General author: 
%%%   \author[number]{Full name}{}
\author[1,2]{Lei Zhang}{}
\author[3,4]{Alexander Wolszczan}{}
\author[5]{Joshua Pritchard}{}
\author[6]{Ryan S. Lynch}{}
\author[7,1]{Di Li}{{dili@mail.tsinghua.edu.cn}}
\author[1,8]{\\Erbil G\"{u}gercino\u{g}lu}{}
\author[1]{Pei Wang}{}
\author[5]{Andrew Zic}{}
\author[2]{Yuanming Wang}{}
\author[2]{Pavan A. Uttarkar}{}
\author[5]{Shi Dai}{}

%\author[1,2]{\\Lei Zhang\footnote{Corresponding author. Email: leizhang996@nao.cas.cn}}{}
%%%\address[number]{Address, City {\rm Postcode}, Country}
\address[1]{State Key Laboratory of Radio Astronomy and Technology, National Astronomical Observatories, Chinese Academy of Sciences, Beijing {\rm 100101}, China}
\address[2]{Centre for Astrophysics and Supercomputing, Swinburne University of Technology, P.O. Box 218, Hawthorn, VIC {\rm 3122}, Australia}
\address[3]{Center for Exoplanets and Habitable Worlds, Pennsylvania State University, 525 Davey Laboratory, University Park, PA 16802, USA}
\address[4]{Department of Astronomy and Astrophysics, Pennsylvania State University, 525 Davey Laboratory, University Park, PA 16802, USA}
\address[5]{CSIRO Space and Astronomy, Australia Telescope National Facility, PO Box 76, Epping, NSW {\rm 1710}, Australia}
\address[6]{Green Bank Observatory, Green Bank, WV {\rm 24401}, USA}
\address[7]{New Cornerstone Science Laboratory, Department of Astronomy, Tsinghua University, Beijing {\rm 100084}, China}
\address[8]{School of Arts and Science, Qingdao Binhai University, Huangdao District, {\rm 266525}, Qingdao, China}

\abstract{
Coherent, periodic radio emission from pulsars has been widely interpreted as evidence of neutron stars as strongly magnetized compact objects. In recent years, radio pulses have also been detected from white dwarfs (WDs) in tight binary systems, raising the question of whether isolated WDs could similarly host pulsar-like emission. We conducted the most sensitive search to date for coherent radio signals from five isolated, rapidly rotating, and magnetized WDs, using the Five-hundred-meter Aperture Spherical radio Telescope (FAST), the Green Bank Telescope (GBT), and the Australia Telescope Compact Array (ATCA). No pulsed or continuum radio emission was detected down to $\mu$Jy levels. These non-detections place the most stringent observational constraints yet on the existence of isolated WD pulsars. Our results suggest that either such emission is intrinsically weak, narrowly beamed, or requires binary-induced magnetospheric interactions absent in solitary systems. Comparison with the known radio-emitting WDs highlights the critical role of companion interaction in enabling detectable emission. This work expands on prior surveys by targeting sources with the most favorable physical conditions for WD pulsar-like activity and employing highly sensitive, targeted observations. Future observations with next-generation facilities such as the SKA will be essential to explore fainter or sporadic emission from massive, magnetic WDs and to investigate their potential as compact radio transients further. 
}

\maketitle

%%%%%%%%%%%%%%%%%%%%%%%%%%%%%%%%%%%%%%%%%%%%%%%%%%%%%%%
%%% The main text. 
%\twocolumn\onecolumn
\begin{multicols}{2}
%%%%%%%%%%%%%%%%%%%%%%%%%%%%%%%%%%%%%%%%%%%%%%%%%%%%%%%
\section{Introduction} \label{sec:intro}
White dwarfs (WDs), the compact remnants of low- to intermediate-mass stars, represent a well-established endpoint of stellar evolution~\cite{Chandrasekhar1939, Schwarzschild1958}. Although typically regarded as inert, slowly cooling objects, a subset of WDs exhibit rapid rotation and strong magnetic fields, suggesting the potential for active magnetospheric processes~\citep{Ferrario1997, Kawaler2015}. Theoretical models predict that, under appropriate conditions, such magnetized WDs could produce coherent radio emission through mechanisms such as curvature-induced pair cascades~\cite{Zhang2005} or electron cyclotron maser instability~\citep{Qu2025}. Additionally, Kashiyama et al.~\cite{Kashiyama2013} proposed that fast radio bursts (FRBs) could originate from double-degenerate WD mergers, wherein coherent radio bursts are emitted from the polar caps of a rapidly rotating, magnetized WD remnant. These models collectively support the plausibility of pulsar-like activity in WDs.

To date, only two confirmed WDs, ARScorpii~\cite{Marsh2016} and J191213.72$-$441045.1~\cite{Pelisoli2023}, have been observed to emit pulsed radio signals with short periodicities (minutes). Both reside in close binary systems with M-dwarf companions, and their radio emission is believed to originate from electrodynamic interactions with the companion, rather than from intrinsic pulsar-like mechanisms within the white dwarf itself~\citep[e.g.,][]{Katz2017, Lyutikov2020, Pelisoli2022, Pelisoli2024}. Additional systems, such as the long-period transients (LPTs) ILT~J1101+5521~\cite{deRuiter2025} and GLEAM-XJ0704$-$37~\cite{HurleyWalker2024, Rodriguez2025}, have also been proposed as WD systems in binary orbits with M-dwarf companions, as confirmed by optical spectroscopy. These systems exhibit much longer modulation periods (hours) and are likewise thought to be powered by binary interaction. Taken together, these findings suggest that companion-induced processes may play a crucial role in enabling detectable radio emission from white dwarfs. In contrast, no pulsed or continuum radio emission has yet been conclusively detected from isolated WDs~\cite{Pelisoli2024, Route2024}.

This motivates a fundamental question: Can magnetized, rapidly rotating isolated WDs generate coherent radio emission on their own? If not, what physical mechanisms suppress emission in the absence of a companion? Addressing these questions is crucial for evaluating the role of WDs in the broader family of compact, magnetized stellar remnants~\citep{Kashiyama2013}, and for testing theories of coherent radio emission across different stellar endpoints.

The potential for pulsar-like emission arises from shared physical traits between isolated WDs and neutron stars, notably rapid rotation and strong surface magnetic fields~\cite{Wickramasinghe2000}. Observational surveys show that only a small fraction of isolated WDs exhibit magnetic fields exceeding $10^9$\,G, and even fewer also have short spin periods ($P < 1000$\,s)~\cite{ferrario15, Hernandez2024}. A notable example is SDSS~J221141.80$+$113604.5 (henceforth WD~2211+113), which exhibits a strong surface magnetic field of $B \sim 15$\,MG and a rapid spin period of 70\,s~\cite{Kilic2021}. This parameter space overlaps with that of PSR~J0901$-$4046, an ultra-long-period radio pulsar with $P = 76$\,s and $B \sim 1.3 \times 10^{14}$\,G~\cite{Caleb2022}, both of which are theoretically capable of supporting coherent emission. This overlap raises the intriguing possibility that some WDs, such as WD~2211+113, may also function as radio pulsars under favorable conditions.

Previous observational efforts have provided only modest constraints. Pelisoli et al.~\cite{Pelisoli2024} searched for radio emission from WDs using Very Large Array Sky Survey (VLASS) data at 8.9 and 11.0\,GHz, reporting non-detections with upper limits of 1–3\,mJy. Route~\cite{Route2024} conducted targeted observations with the Arecibo telescope at 5\,GHz, though without a specific focus on pulsed radio emission. These studies underscore the challenges of detecting weak radio emission from WDs and suggest the need for deeper, lower-frequency observations with improved sensitivity.

In this study, we present the most sensitive targeted search to date for radio emission from isolated WDs. We selected five sources with both rapid rotation and strong magnetic fields, and achieved sensitivity levels down to the $\mu$Jy regime using  Five-hundred-meter Aperture Spherical radio Telescope (FAST; \cite{Nan2011, Li2018}), the Green Bank Telescope (GBT), and the Australia Telescope Compact Array (ATCA) observations. Section~\ref{sec:obsre} describes the observations and results, Section~\ref{sec:dis} discusses their implications, and Section~\ref{sec:sum} summarizes our conclusions.

\section{Observations and Results} \label{sec:obsre}
\begin{table*}
\centering
\caption{ List of isolated WDs observed in this work and the two known pulsar-like WDs in binary systems exhibiting pulsed radio emission (periods $< 1000$\,s)}, along with their basic parameters from the literature.
\setlength{\tabcolsep}{0.55mm}{
\label{tab:WDs_inf}
\begin{tabular}{lcccccccc}\hline
Name                       & RAJ         & DecJ         & Period   & Mass        & Magnetic field    & Distance  & Orbital period     & Ref \\
                           & (hh:mm:ss)  & (dd:mm:ss)   & (s)      & M$_{\odot}$ & (MG)              & (pc)      & (hour)     &     \\\hline
\multicolumn{8}{c}{\textbf{Isolated WDs}}  \\ \hline
WD 0316-849                & 03:17:15.84 & -85:32:25.55 & 725.4    & 1.26        & 450         & 29.38 $\pm$ 0.02  & -- & \cite{Brinkworth2004, Harayama2013}\\
SDSS J125230.93$-$023417.7 & 12:52:30.93 & -02:34:17.72 & 317.278  & 0.56        & 5           & 77.1 $\pm$ 0.7    & -- & \cite{Reding2020}\\
WD 1832+089                & 18:32:02.82 & +08:56:36.24 & 353.456  & 1.33        & $\lesssim$1 & 75.55 $\pm$ 0.55  & -- & \cite{Pshirkov2020}\\
WD 1859+148                & 19:01:32.73 & +14:58:07.17 & 416.242  & 1.32        & 600-900     & 41.40 $\pm$ 0.08  & -- & \cite{Caiazzo2021}\\
WD 2211+113                & 22:11:41.80 & +11:36:04.64 & 70.32    & 1.27        & 15          & 68.86 $\pm$ 1.54  & -- & \cite{Kilic2021}\\\hline
\multicolumn{8}{c}{\textbf{Radio-pulsing WDs in binary}}  \\ \hline
AR Scorpii                 & 16:21:47.28 & -22:53:10.38 & 117      & 0.39        & 100-500     & 116 $\pm$ 16   & 3.56 & \cite{Marsh2016, Buckley2017, Katz2017}\\
J191213.72$-$441045.1      & 19:12:13.72 & -44:10:45.10 & 318      & 0.59        & 50          & 237 $\pm$ 5    & 4.03 & \cite{Pelisoli2023, Pelisoli2024}\\\hline
\end{tabular}}
\end{table*}

To investigate potential radio emission from rapidly rotating isolated WDs, we selected five candidates with rotation periods below 1000\,s~\cite{Hernandez2024}, as listed in Table~\ref{tab:WDs_inf}. These targets were primarily chosen for their combination of strong magnetic fields and rapid rotation—key characteristics expected to favor the generation of coherent radio emission. One exception is WD~1832+089, which, despite having only an upper limit on its surface magnetic field ($B \lesssim 1$\,MG), was included due to its exceptionally short spin period ($P = 416$\,s). This object probes the regime of fast rotators near the boundary of the parameter space predicted to support pulsar-like activity.

\begin{figure*}[!]
    \centering
    \includegraphics[width=0.7\linewidth]{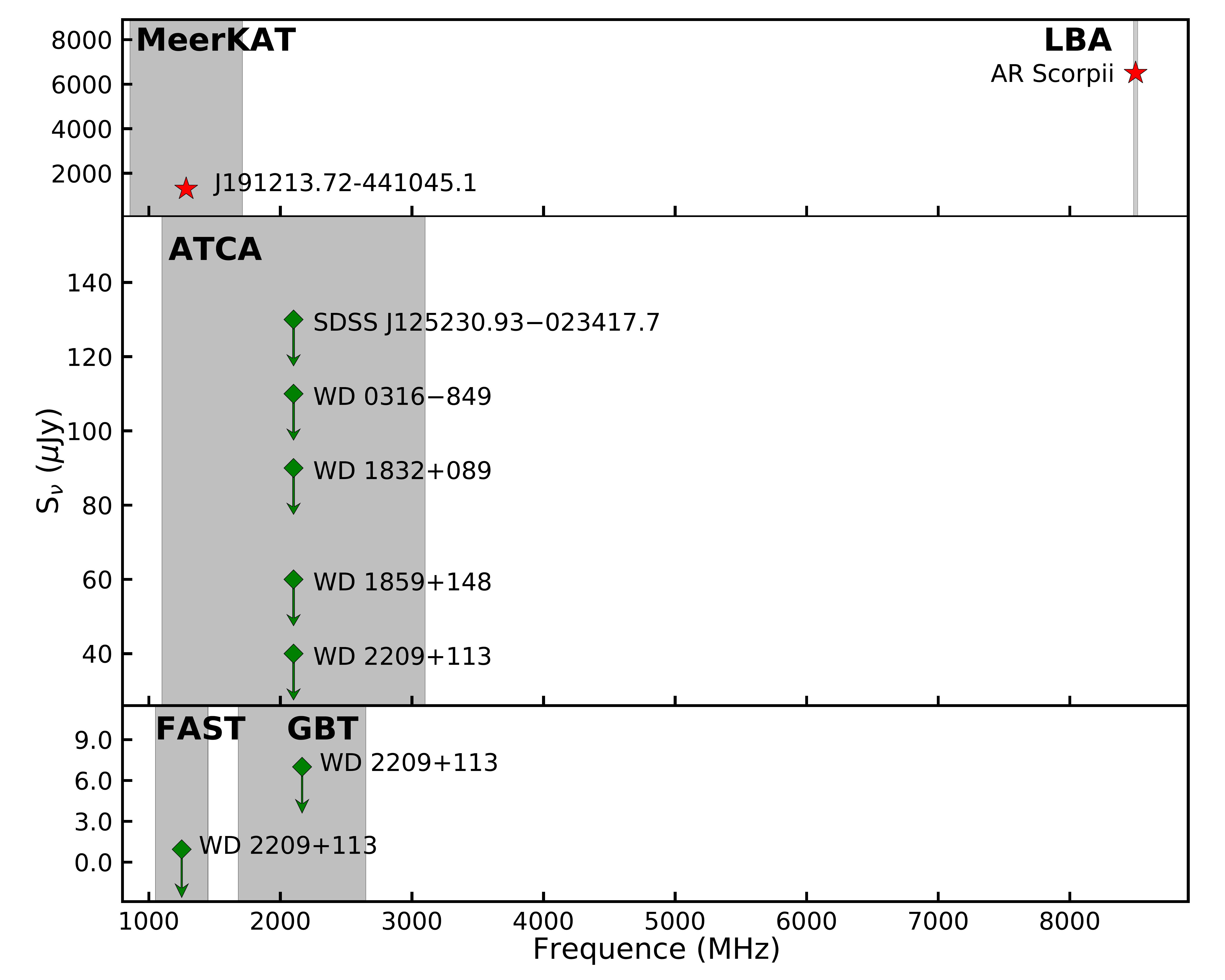}
    \caption{Average flux densities of seven WDs as a function of observing frequency. Green diamonds with downward arrows indicate upper limits for five isolated WDs from this work, based on observations with FAST, GBT, and ATCA. Red stars denote the two known pulsar-like WDs in binary systems exhibiting pulsed radio emission (periods $<$ 1000 s)~\cite{Marsh2016,Marcote2017,Pelisoli2023,deRuiter2025}. Gray shaded regions represent the frequency coverage of the respective telescopes.}
    \label{fig:flux-freq}
\end{figure*}

A review of archival radio survey data—including the Rapid ASKAP Continuum Survey (RACS at 887.5, 943.5, 1367.5, and 1655.5 MHz), the ASKAP Variables and Slow Transients Survey (VAST; 887.5 MHz), and the NRAO VLA Sky Survey (NVSS; 1400 MHz)—revealed no significant radio detections at the positions of our selected targets. To probe deeper radio sensitivity, we subsequently conducted targeted observations with FAST, GBT, and ATCA. The resulting flux density upper limits for the five isolated WDs observed in this study are shown in Figure~\ref{fig:flux-freq}, based on these dedicated observations.

\subsection{FAST}
Using the FAST 19-beam receiver at L-band, we conducted an in-depth search for pulsed radio emission from WD 2211+113, the most rapidly rotating isolated WD, on 24 September 2023 through the observing project PT2023\_0206. The observation was performed with the central beam, which has a beam width of 3 arcminutes at 1250 MHz, and an integration time of 80 minutes. Data collection employed 8-bit sampling in pulsar search mode. The observed frequency range spanned 1000 MHz to 1500 MHz, divided into 4096 frequency channels. However, due to bandpass roll-off, the effective bandwidth was limited to 1050 MHz to 1450 MHz. The sampling time was 49.152\,$\upmu$s.

The observation data were processed to remove radio frequency interference (RFI) and subsequently de-dispersed using a range of trial dispersion measures (DMs) between 0 and 30\,pc cm$^{-3}$, covering the estimated DM values from Galactic electron density models (YMW16~\cite{Yao2017} = 0.72\,pc cm$^{-3}$ and NE2001~\cite{Cordes2002} = 0.34\,pc cm$^{-3}$), considering the source position and its inferred distance. 

For each de-dispersed time series, we conducted a blind Fourier-domain periodicity search using a \textsc{presto}-based pipeline~\citep{Ransom2002}, optimized to detect fast-spinning pulsars, including those in compact binary systems. Although WD~2211+113 is an isolated white dwarf with no known companion, the 80-minute FAST observation was subdivided into 20- and 40-minute segments to enhance sensitivity to any serendipitous pulsars within the telescope beam, particularly those in short-period binary orbits~\cite{Zhang2023}. This strategy allows the detection of binary pulsars with orbital periods as short as $\sim$3 hours~\cite{Ng2015}, ensuring a comprehensive search for unrelated sources within the field of view.

Searching for long-period signals with single-dish telescopes presents significant challenges, primarily due to red noise arising from system temperature variations, gain fluctuations, and sky background changes over extended integration times~\citep{Lazarus2015, Singh2022}, as well as the effects of pulse nulling~\citep{Backer1970, Zhang2019}, which reduce the sensitivity of standard Fourier-based techniques. To mitigate these limitations, we employed the Fast Folding Algorithm (FFA), which is better suited for detecting long-period pulsars~\citep{Morello2020, Zhou2024}. A targeted FFA search was performed using a \textsc{riptide}-based pipeline~\citep{Morello2020}, optimized for long-period signals. The search was restricted to a narrow window of $\pm$10 s around the known spin period of WD~2211+113 (70.32 s; \cite{Kilic2021}), and red noise mitigation was applied within the pipeline. The time resolution was adjusted to retain sensitivity to narrow duty-cycle pulses. No candidates were consistently recovered across multiple trials, suggesting that any apparent detections were likely due to statistical noise.

A separate search was conducted to detect potential single pulses using a \textsc{heimdall}~\cite{Barsdell2012}-based pipeline, aiming to identify transient signals similar to Rotating Radio Transients (RRATs) and Fast Radio Bursts (FRBs). The trial DM range extended up to 5,000\,pc cm$^{-3}$, and a boxcar search was applied for pulse widths up to 60 ms. However, no significant single pulses were identified.

The minimum detectable flux density for a targeted pulsar search can be estimated using the radiometer equation~\citep{Lorimer2004}:
\begin{equation} 
S_{min}= \beta\frac{(S/N_{min}) T_{\text{sys}}}{G\sqrt{n_{p}t_{\text{int}}\Delta f}}\sqrt{\frac{W}{P-W}}, 
\end{equation}
where the sampling efficiency $\beta$ is set to 1 for our FAST 8-bit recording system. We adopt a minimum signal-to-noise threshold of $S/N_{min} = 10$ for detection. The system temperature, $T_{\text{sys}}$ is $\simeq$ 24 K\footnote{$t_{\text{sys}}$ includes the receiver temperature as well as contributions from the sky, atmosphere, and ground spillover.}, while the antenna gain ($G$) for the central beam is $\simeq$ 16 K Jy$^{-1}$~\cite{Jiang2020}. The number of polarizations ($n_{p}$) is 2, and the integration time ($t_{\text{int}}$) is the integration time in units of seconds. The effective bandwidth ($\Delta f$) is 350 MHz, after accounting for the removal of approximately 12.5\% of frequency channels affected by radio frequency interference (RFI). For the case of a 70.32 s spin period considered in this work, we adopt a 1\% duty cycle\footnote{Adopting a broader duty cycle (e.g., 5–10\%, as seen in typical pulsars~\cite{Zhang2025}) would increase the flux density threshold but does not alter our primary conclusions.}, corresponding to a pulse width of $W \simeq 0.7$ s. This assumption is motivated by the ultra-long-period pulsar PSR~J0901$-$4046, which has a spin period of 76 s and exhibits L-band pulse widths of approximately 1\% of its period~\cite{Caleb2022}.

For an FFA search, in principle, has a correction factor of 1/0.93~\cite{Morello2020}. For the practical case of FAST data, we follow the empirical approach of Zhou et al.~\cite{Zhou2024}, who conducted pulse injection simulations into actual FAST observations to calibrate detection thresholds. We injected synthetic pulsars with $P = 70$ s and a 1\% duty cycle into our data, and confirmed that the practical detection threshold is approximately 15\% higher than predicted by the standard radiometer equation. Incorporating these instrumental and observational parameters, we estimate the minimum detectable flux density for our FAST L-band observation to be approximately 0.95 $\upmu$Jy.
%%% FAST parameters
%SN = 10 
%W = 0.01     # duty cycle
%Tsys = 24    # K
%G = 16       # K/Jy the gain the centralbeam at 1250 MHz
%BW = 350     # effective bandwidth
%Tobs = 80*60    # s
%flux = (SN*Tsys/G) * 1/np.sqrt(2*BW*Tobs) * np.sqrt(W/(1-W)) * 1000
%print("FAST_estimate: If SN = %.0f, flux(uJy) = %.5f" %(SN,flux))

\subsection{GBT}
We observed WD2211+1136 using the Robert C.\ Byrd Green Bank Telescope as part of project AGBT22B-336 on 2024 January 8 (MJD 63017.926425). We used the S-Band receiver centered at a frequency of 2165 MHz with a nominal total bandwidth of 970 MHz, with frequency channels approximately 0.732 MHz wide.  However, radio frequency interference reduced the usable bandwidth to approximately 865 MHz. We observed for a total of 1791.5 seconds and sampled data every 21.845 $\mu$s.  We recorded polarization self and cross terms but in the analysis presented here we summed polarizations and worked only with total intensity data.

We obtained a parallax-derived distance of 68.9 pc from the Gaia Archive~\cite{Gaia2016,Gaia2023}.  At this distance, the NE2001 model~\cite{Cordes2002} predicts a dispersion measure (DM) of only 0.34 pc cm$^{-3}$, while the YMW16 model predicts 0.72 pc cm$^{-3}$~\cite{Yao2017}. At observing frequencies around 2 GHz such a low DM will lead to only a 0.3--0.6 ms delay between the top and bottom of the observing band, which is insignificant when compared to the 70.3-s rotational period of the WD.  As such, we do not expect dispersive delay to limit our sensitivity to pulsed emission.  Nevertheless, we dedispersed our data at the NE2001 DM of 0.34 pc cm$^{-3}$ for subsequent searches.

We searched for periodic and intermittent pulses using the \texttt{PRESTO} software package~\cite{Ransom2002}.  We removed RFI using the \texttt{rfifind} routine and performed a Fourier-domain search of the dedispersed time series using \texttt{accelsearch}.  We searched for bright, single pulses using the \texttt{single\_pulse\_search.py} routine. Neither method resulted in significant candidates.  We also used the \texttt{rednoise} routine to reduce noise at low Fourier frequencies and repeated our search, but still not did not find any candidates.  Finally, we used the \texttt{dspsr}\footnote{\url{https://dspsr.sourceforge.net/}} software package to fold the data at the nominal rotational period and DM of the WD, and used tools from the \texttt{PSRCHIVE} package \citep{Hotan2004} to calibrate the data and remove RFI.  No pulsed emission was visible.  From our flux-calibrated data we estimate a 1-$\sigma$ flux density limit of 7 $\mu$Jy.\\\\\\

\subsection{ATCA}
\begin{table*}
\centering
\caption{Summary of ATCA observations used in this work, including $1\sigma$ flux density limits derived from: the full synthesis image ($S_\nu$, Image), a dynamic spectrum binned at 1-minute intervals and 16 MHz bandwidth ($S_\nu$, DS), a 1-minute resolution light curve ($S_\nu$, LC), a phase-folded dynamic spectrum with 16 MHz frequency and phase bins spanning the full rotation period ($S_\nu$, Folded DS), and a phase-folded light curve ($S_\nu$, Folded LC).}
\label{tab:ATCA_obs}
\renewcommand{\arraystretch}{1.0}
\setlength{\tabcolsep}{1.0mm}{
\begin{tabular}{cccccccc}\hline 
Observation & Observation & $S_\nu$        & $S_\nu$    & $S_\nu$   & $S_\nu$     & $S_\nu$     & Configurations \\
Date        & Length      & (image)        & (DS)       & (LC)      & (Folded DS) & (Folded LC) &  \\
(UTC)       &  (hr)       & ($\mu$Jy/beam) & ($\mu$Jy)  & ($\mu$Jy) & ($\mu$Jy)   &  $\mu$Jy    &   \\\hline 
\multicolumn{3}{l}{\textbf{WD~0316$-$849}}\\
2024-10-12  &  4.3 & 30   & 4700  & 500  & 1200 & 140 & 6A \\
2025-01-18  &  4.0 & 40   & 5500  & 660  & 1500 & 180 &  6C \\
2025-03-09  &  5.3 & 20   & 3800  & 540  & 910  & 110 &  6D \\  
\multicolumn{3}{l}{\textbf{SDSS J125230.93$-$023417.7}}\\            
2024-10-21  &  4.3 & 90   & 3600  & 1090 & 780  & 130 & 6A \\
2025-01-18  &  1.3 & 1100 & 12100 & 1090 & 3810 & 150 & 6C \\
2025-02-23  &  4.8 & 270  & 6300  & 3080 & 1680 & 530 & 6D \\
\multicolumn{3}{l}{\textbf{WD~1832+089 }}\\
2024-10-19  &  4.0 & 90   & 5200  & 700  & 1510 & 210 & 6A \\
2025-01-05  &  4.1 & 180  & 4500  & 630  & 1180 & 90  & 6C \\
2025-03-07  &  3.9 & 70   & 3900  & 480  & 1030 & 100  & 6D \\
\multicolumn{3}{l}{\textbf{WD~1859+148 }}\\
2024-10-13  &  3.9 & 30   & 2100  & 230  & 520  & 60  & 6A \\
2025-01-10  &  4.1 & 25   & 3500  & 420  & 860  & 60  & 6C \\
2025-02-28  &  4.0 & 30   & 2700  & 330  & 870  & 150 & 6D \\
\multicolumn{3}{l}{\textbf{WD~2211+113}}\\
2024-10-20  &  4.5 & 20   & 1980  & 290  & 370  & 50  & 6A \\
2025-01-19  &  4.3 & 30   & 2820  & 340  & 520  & 40  & 6C \\
2025-03-08  &  4.2 & 30   & 2940  & 340  & 540  & 80  & 6D \\\hline         
\end{tabular}}
\end{table*}
In parallel, we carried out three deep continuum imaging observations for each of the five selected WDs using ATCA, with a central observing frequency at 2.1 GHz and a bandwidth of 2 GHz (Project C3656). The primary objective of these observations was to identify persistent radio emission and search for pulsed emission at the WD rotation period. We calibrated the bandpass response and flux scale with PKS~1934$–$638 and corrected for time-varying gains in each observation using interleaved scans of a nearby gain calibrator. We used standard continuum data reduction routines with {\sc miriad} \citep{Sault1995} to flag and calibrate the data, and used {\sc WSClean} \citep{Offringa2014} to image the full primary beam and produce a model of all sources of emission in the field. We used the \textsc{DStools}\footnote{\url{https://github.com/askap-vast/dstools}} \citep{Pritchard2024b} package to subtract the field model from the visibilities, baseline average the residual visibilities phased towards our target, and form dynamic spectra binned to 1-minute / 16 MHz resolution to search for pulsed emission in the time-frequency plane. We further searched for rotationally modulated pulsed emission by folding dynamic spectra and lightcurves to the rotation period of each WD. We detect no evidence of pulsed emission or continuum emission in any of the 15 observations, and provide upper limits on the measured flux density in each of these searches in Table~\ref{tab:ATCA_obs}.

\section{Discussion} \label{sec:dis}
\subsection{Interpretation of Radio Non-Detections}
Our deep radio observations using FAST, GBT, and ATCA have revealed no significant detection of either pulsed or persistent radio emission from any of the five isolated WDs observed in this study (listed in Table~\ref{tab:WDs_inf}). These non-detections, reaching sub-microjansky sensitivity levels, impose meaningful upper limits on radio luminosity and place important constraints on theoretical models of WD magnetospheres.

Several factors could account for the absence of detectable emission. One possibility is unfavorable beaming geometry: if the emission is highly directional and the beam does not intersect our line of sight, detection is inherently unlikely---an issue well documented in studies of neutron star pulsars. Another consideration is that coherent emission, if present, may simply be too weak to detect with current instrumentation. While models have proposed that WDs with rapid rotation and strong magnetic fields could generate radio emission via mechanisms such as electron cyclotron maser instability or curvature radiation~\cite{Qu2025}, the conditions required---strong accelerating potentials and efficient pair production---may not be met in isolated systems.

Magnetic isolated WDs spin down slower compared to neutron stars due to their larger moments of inertia, longer spin periods, and lower surface magnetic fields. Since the magnetic fields of WDs would not be subjected to appreciable field decay due to longer Ohmic dissipation timescales \cite{cumming02}, current isolated WD periods and magnetic field strengths should closely reflect those at their birth and therefore can be taken as constants. Under these assumptions, the minimum surface magnetic field strength that a WD should have to ignite radio emission can be recast with scaling from neutron star parameters as \cite{zhang00}
\begin{equation}
    B_{\rm s}\gtrsim2.8\times10^{9}\mbox{G}\left(\frac{R_{\rm WD}}{6000\,\mbox{km}}\right)^{-19/8}\left(\frac{P_{\rm s}}{100\,\mbox{s}}\right)^{15/8},
    \label{puredip}
\end{equation}
for pure dipolar magnetic field configuration and
\begin{equation}
    B_{\rm s}\gtrsim2.3\times10^{8}\mbox{G}b^{-1/4}\left(\frac{R_{\rm WD}}{6000\,\mbox{km}}\right)^{-2}\left(\frac{P_{\rm s}}{100\,\mbox{s}}\right)^{3/2},
    \label{muldip}
\end{equation}
for twisted and multipolar, Sunspot-like magnetic field configuration. In the above expressions, $R_{\rm WD}$ is the WD radius, $P_{\rm s}$ is the spin period of the WD and $b\lesssim10$ is the ratio between the spot's magnetic field and the dipolar strength. All 37 observed isolated magnetic WDs with measured surface magnetic fields listed in \citep{ferrario20} fall well below the radio emission death lines determined by Eqs. (\ref{puredip}) and (\ref{muldip}) \cite{rea24}, possibly explaining their non-detection in the radio band. Note, however, that for currently known isolated magnetic WDs with reliable measurements of spin periods and surface magnetic fields, the relation between these two quantities does not exhibit a clear correlation like those predicted by Eqs. (\ref{puredip}) and (\ref{muldip}) \cite{ferrario15}, implying that magnetic field generation mechanism may differ among different objects, including possibilities of flux-freezing during the collapse of the progenitor star \cite{reisenegger09}, dynamo processes during formation through merger \cite{tout08} and convective processes during the crystallization epoch of the liquid core at later times of the WD evolution \cite{isern17}.

Obtaining a long-lived, highly-ordered and collimated radiation with a small duty cycle from a WD faces with yet another theoretical difficulty \cite{beniamini23}. In order to achieve a high degree of beaming, a bunch of relativistic particles with a large Lorentz factor is usually required. However, the physical conditions in the vicinity of the surface of a WD, unlike a neutron star, are not sufficient to source and eject plenty of charged particles, and accelerate plasma to relativistic speeds. Furthermore, photon and magnetic field involved pair production cascades entail intense magnetic fields and small curvature radii \cite{wadiasingh19}. Whereas a neutron star can maintain the required twist by means of plastic flow in its crust \cite{suvorov23}, internal hydromagnetic stresses associated with superconducting core \cite{thompson01} or magnetospheric current coupling \cite{akgün16}, such mechanisms are lacking for WDs.

In particular, the large radii and relatively weak magnetic fields of WDs compared to neutron stars result in lower electric fields, reducing the likelihood of magnetospheric particle acceleration. Although two-photon pair production has been suggested as a viable channel for plasma generation in WDs \citep{Philippov2018}, this process depends on sufficient photon densities, which may not be present without a companion star or other external sources. Our results indicate that isolated WDs may lack the necessary conditions to form the plasma-rich magnetospheres required to sustain coherent radio emission.

\subsection{Comparison with Known Radio-Emitting WDs in Binaries}
To date, all confirmed WDs known to exhibit detectable radio emission are found in close binary systems with M-dwarf companions~\cite{Marsh2016, Pelisoli2023, deRuiter2025, Rodriguez2025}. In these systems, the observed radio emission is not attributed to intrinsic pulsar-like processes in the WD itself, but rather to magnetospheric interactions between the WD and its companion. Such interactions can induce electrical currents or magnetic reconnection events, producing broadband emission that is periodically modulated, typically at the spin or orbital period of the system.

In AR Scorpii, for example, the observed pulsed emission is strongly modulated at the WD’s 117\,s spin period and varies with orbital phase, supporting a model in which the emission is powered by electrodynamic coupling with the companion star \citep{Takata2017, Katz2017}. These characteristics differ fundamentally from the isolated systems we observed, where no such interaction is present.

This comparison underscores the importance of binary interaction in enabling detectable radio emission from WDs. It also suggests that the emission mechanisms active in systems like AR Scorpii may not be generalizable to the broader population of isolated WDs, which would rely solely on spin-down power to energize their magnetospheres.

\subsection{Constraints on the WD Pulsar Hypothesis}
The long-standing hypothesis that some isolated WDs could function as radio pulsars—powered by rotation and magnetospheric particle acceleration—has not yet been confirmed observationally~\citep{Zhang2005}. Our study, targeting five of the most promising candidates selected for their short spin periods and strong magnetic fields, places the most stringent observational constraints to date on this scenario.

In particular, WD~2211+113, the fastest known isolated WD, with a spin period of 70.32 s and a high inferred magnetic field, showed no evidence of pulsed or persistent radio emission in our deep observations. If this object were emitting coherent radio waves via a pulsar-like mechanism, even with modest efficiency, such emission would likely have been detectable. One possibility for the non-detection is that our line of sight does not intersect the emission beam.

Given the null results for all five isolated WDs, we can place an empirical upper limit on the beaming fraction. Assuming randomly oriented beams, the probability of detecting none of five emitting WDs implies a beaming fraction $f \lesssim 1/5 = 20\%$. For comparison, typical beaming fractions for neutron star radio pulsars are around 10\%, depending on the spin period, pulse width, and magnetic inclination~\cite{Tauris1998, Kolonko2004}. This suggests that geometric effects alone could plausibly account for our non-detections, particularly if WD emission cones are similarly narrow or more tightly collimated.

Our findings argue against the existence of a bright, rotation-powered population of WD pulsars analogous to neutron star pulsars. While we cannot exclude the possibility of rare systems with favourable geometries or enhanced emission efficiencies, such sources are intrinsically rare or radiate below the sensitivity limits of current instruments. Continued searches with improved sensitivity, broader frequency coverage, and longer integration times will be essential to further probe this hypothesis.

\section{Summary and Conclusions} \label{sec:sum}
In this study, we conducted a targeted search for coherent radio emission from five rapidly rotating, magnetized isolated WDs using the FAST, GBT, and ATCA. This investigation addresses a key open question in compact object astrophysics: whether isolated WDs can generate detectable non-thermal radio emission through mechanisms analogous to those operating in neutron star pulsars.

Despite reaching $\upmu$Jy sensitivities in both pulsed and continuum regimes, we detected no significant emission from any of the targets. These non-detections place stringent upper limits on the radio luminosity of isolated WDs, suggesting that either coherent radio emission from such systems is intrinsically weak, highly beamed away from Earth, or that the necessary magnetospheric conditions for particle acceleration and plasma generation are not realized in isolated WDs.

Comparison with known radio-emitting confirmed WDs~\cite{Marsh2016, Pelisoli2023, deRuiter2025, Rodriguez2025} indicates that binary interaction plays a critical role in producing observable radio emission. In contrast, isolated WDs likely lack the external triggers necessary to energize their magnetospheres and drive strong radio emission.

Our results provide the most stringent targeted observational constraints to date on the WD pulsar hypothesis. While our null detections suggest that a luminous, rotation-powered population of isolated WD pulsars is either rare or has a low beaming fraction, they do not rule out the existence of fainter or sporadically active emitters. Future discoveries may still be possible with next-generation instruments that offer improved sensitivity and broader frequency coverage.

Looking ahead, upcoming facilities such as the Square Kilometre Array (SKA) and the Deep Synoptic Array (DSA-2000) will be crucial in probing this elusive population. Extending search efforts to lower frequencies, longer integration times, and broader samples of massive, magnetic WDs will improve the prospects of uncovering coherent radio emission from isolated WDs, thereby providing new insights into the physics of compact magnetized remnants.

%%%%%%%%%%%%%%%%%%%%%%%%%%%%%%%%%%%%%%%%%%%%%%%%%%%%%%%
%%% Acknowledgements. 
%%%%%%%%%%%%%%%%%%%%%%%%%%%%%%%%%%%%%%%%%%%%%%%%%%%%%%%
\subsection*{\begin{center}ACKNOWLEDGMENTS\end{center}}
This work was supported by the National Natural Science Foundation of China (NSFC; grant Nos. 12588202, 12103069, 11725313, 12273008, 12373109, and 12203045), CAS project No. JZHKYPT-2021-06, and the Open Project Program of the Key Laboratory of FAST, Chinese Academy of Sciences. Di Li is a New Cornerstone Investigator.
The Australia Telescope Compact Array (ATCA) is part of the Australia Telescope National Facility managed by CSIRO. This research also utilized data from the Australian Square Kilometre Array Pathfinder (ASKAP), accessed via the CASDA data archive (\url{https://research.csiro.au/casda/casda-user-guide/#14_Acknowledging_ASKAP_data_and_CASDA_in_publications}).
This material is based upon work supported by the National Radio Astronomy Observatory and the Green Bank Observatory, which are major facilities funded by the U.S. National Science Foundation and operated by Associated Universities, Inc.
We acknowledge use of data from the ESA \textit{Gaia} mission (\url{https://www.cosmos.esa.int/gaia}), processed by the DPAC (\url{https://www.cosmos.esa.int/web/gaia/dpac/consortium}), with funding provided by national institutions participating in the \textit{Gaia} Multilateral Agreement.\\\\
%%%%%%%%%%%%%%%%%%%%%%%%%%%%%%%%%%%%%%%%%%%%%%%%%%%%%%%
%%% Conflict of interest. 
%%%%%%%%%%%%%%%%%%%%%%%%%%%%%%%%%%%%%%%%%%%%%%%%%%%%%%%
%\InterestConflict{The authors declare that they have no conflict of interest.}
{\bf Conflict of interest} The authors declare that they have no conflict of interest.

%%%%%%%%%%%%%%%%%%%%%%%%%%%%%%%%%%%%%%%%%%%%%%%%%%%%%%%
%%% Reference section. 
%%% citation in the content using "some words~\cite{1,2}".
%%% ~ is needed to make the reference number is on the same line with the word before it.
%%%%%%%%%%%%%%%%%%%%%%%%%%%%%%%%%%%%%%%%%%%%%%%%%%%%%%%
\clearpage
\bibliographystyle{scpma-zycai} %Citation order, maximum 50 authors
\bibliography{ms}

\begin{thebibliography}{10}
\providecommand{\url}[1]{\texttt{#1}}
\providecommand{\urlprefix}{URL }
\providecommand{\doi}[1]{doi:~\href{http://doi.org/#1}{\nolinkurl{#1}}}
\providecommand{\arXiv}[1]{\href{https://arxiv.org/abs/#1}{\nolinkurl{https://arxiv.org/abs/#1}}}
\providecommand{\eprint}[1]{\href{http://arxiv.org/abs/#1}{\nolinkurl{#1}}}

\bibitem{Chandrasekhar1939}
S.~{Chandrasekhar}, \emph{{An introduction to the study of stellar structure}}
  (1939).

\bibitem{Schwarzschild1958}
M.~{Schwarzschild}, \emph{{Structure and evolution of the stars}} (1958).

\bibitem{Ferrario1997}
L.~{Ferrario}, S.~{Vennes}, D.~T. {Wickramasinghe}, J.~A. {Bailey}, and D.~J.
  {Christian}, \href{http://dx.doi.org/10.1093/mnras/292.2.205}{\mnras}
  \textbf{292}, 205 (1997).

\bibitem{Kawaler2015}
S.~D. {Kawaler}, in \emph{19th European Workshop on White Dwarfs}, (edited by
  P.~{Dufour}, P.~{Bergeron}, and G.~{Fontaine}), volume 493 of
  \emph{Astronomical Society of the Pacific Conference Series}, 65 (2015),
  arXiv: \eprint{1410.6934}.

\bibitem{Zhang2005}
B.~{Zhang} and J.~{Gil}, \href{http://dx.doi.org/10.1086/497428}{\apjl}
  \textbf{631}, L143 (2005), arXiv: \eprint{astro-ph/0508213}.

\bibitem{Qu2025}
Y.~{Qu} and B.~{Zhang}, \href{http://dx.doi.org/10.3847/1538-4357/adb1b5}{\apj}
  \textbf{981}, 34 (2025), arXiv: \eprint{2409.05978}.

\bibitem{Kashiyama2013}
K.~{Kashiyama}, K.~{Ioka}, and P.~{M{\'e}sz{\'a}ros},
  \href{http://dx.doi.org/10.1088/2041-8205/776/2/L39}{\apjl} \textbf{776}, L39
  (2013), arXiv: \eprint{1307.7708}.

\bibitem{Marsh2016}
T.~R. {Marsh}, B.~T. {G{\"a}nsicke}, S.~{H{\"u}mmerich}, F.~J. {Hambsch},
  K.~{Bernhard}, C.~{Lloyd}, E.~{Breedt}, E.~R. {Stanway}, D.~T. {Steeghs},
  S.~G. {Parsons}, et~al., \href{http://dx.doi.org/10.1038/nature18620}{\nat}
  \textbf{537}, 374 (2016), arXiv: \eprint{1607.08265}.

\bibitem{Pelisoli2023}
I.~{Pelisoli}, T.~R. {Marsh}, D.~A.~H. {Buckley}, I.~{Heywood}, S.~B. {Potter},
  A.~{Schwope}, J.~{Brink}, A.~{Standke}, P.~A. {Woudt}, S.~G. {Parsons},
  et~al., \href{http://dx.doi.org/10.1038/s41550-023-01995-x}{Nature Astronomy}
  \textbf{7}, 931 (2023), arXiv: \eprint{2306.09272}.

\bibitem{Katz2017}
J.~I. {Katz}, \href{http://dx.doi.org/10.3847/1538-4357/835/2/150}{\apj}
  \textbf{835}, 150 (2017), arXiv: \eprint{1609.07172}.

\bibitem{Lyutikov2020}
M.~{Lyutikov}, M.~{Barkov}, M.~{Route}, D.~{Balsara}, P.~{Garnavich}, and
  C.~{Littlefield}, \href{http://dx.doi.org/10.48550/arXiv.2004.11474}{arXiv
  e-prints} arXiv:2004.11474 (2020), arXiv: \eprint{2004.11474}.

\bibitem{Pelisoli2022}
I.~{Pelisoli}, T.~R. {Marsh}, S.~G. {Parsons}, A.~{Aungwerojwit}, R.~P.
  {Ashley}, E.~{Breedt}, A.~J. {Brown}, V.~S. {Dhillon}, M.~J. {Dyer}, M.~J.
  {Green}, et~al., \href{http://dx.doi.org/10.1093/mnras/stac2391}{\mnras}
  \textbf{516}, 5052 (2022), arXiv: \eprint{2208.08450}.

\bibitem{Pelisoli2024}
I.~{Pelisoli}, L.~{Chomiuk}, J.~{Strader}, T.~R. {Marsh}, E.~{Aydi}, K.~C.
  {Dage}, R.~{Kyer}, I.~{Molina}, T.~{Panurach}, R.~{Urquhart}, et~al.,
  \href{http://dx.doi.org/10.1093/mnras/stae1275}{\mnras} \textbf{531}, 1805
  (2024), arXiv: \eprint{2402.11015}.

\bibitem{deRuiter2025}
I.~{de Ruiter}, K.~M. {Rajwade}, C.~G. {Bassa}, A.~{Rowlinson}, R.~A.~M.~J.
  {Wijers}, C.~D. {Kilpatrick}, G.~{Stefansson}, J.~R. {Callingham}, J.~W.~T.
  {Hessels}, T.~E. {Clarke}, et~al.,
  \href{http://dx.doi.org/10.1038/s41550-025-02491-0}{Nature Astronomy}
  \textbf{9}, 672 (2025).

\bibitem{HurleyWalker2024}
N.~{Hurley-Walker}, S.~J. {McSweeney}, A.~{Bahramian}, N.~{Rea},
  C.~{Horv{\'a}th}, S.~{Buchner}, A.~{Williams}, B.~W. {Meyers}, J.~{Strader},
  E.~{Aydi}, et~al., \href{http://dx.doi.org/10.3847/2041-8213/ad890e}{\apjl}
  \textbf{976}, L21 (2024), arXiv: \eprint{2408.15757}.

\bibitem{Rodriguez2025}
A.~C. {Rodriguez}, \href{http://dx.doi.org/10.1051/0004-6361/202553684}{\aap}
  \textbf{695}, L8 (2025), arXiv: \eprint{2501.03315}.

\bibitem{Route2024}
M.~{Route}, \href{http://dx.doi.org/10.3847/1538-4357/ad9567}{\apj}
  \textbf{977}, 261 (2024), arXiv: \eprint{2411.13718}.

\bibitem{Wickramasinghe2000}
D.~T. {Wickramasinghe} and L.~{Ferrario},
  \href{http://dx.doi.org/10.1086/316593}{\pasp} \textbf{112}, 873 (2000).

\bibitem{ferrario15}
L.~{Ferrario}, D.~{de Martino}, and B.~T. {G{\"a}nsicke},
  \href{http://dx.doi.org/10.1007/s11214-015-0152-0}{\ssr} \textbf{191}, 111
  (2015), arXiv: \eprint{1504.08072}.

\bibitem{Hernandez2024}
M.~S. {Hernandez}, M.~R. {Schreiber}, J.~D. {Landstreet}, S.~{Bagnulo}, S.~G.
  {Parsons}, M.~{Chavarria}, O.~{Toloza}, and K.~J. {Bell},
  \href{http://dx.doi.org/10.1093/mnras/stae307}{\mnras} \textbf{528}, 6056
  (2024), arXiv: \eprint{2401.15158}.

\bibitem{Kilic2021}
M.~{Kilic}, A.~{Kosakowski}, A.~G. {Moss}, P.~{Bergeron}, and A.~A. {Conly},
  \href{http://dx.doi.org/10.3847/2041-8213/ac3b60}{\apjl} \textbf{923}, L6
  (2021), arXiv: \eprint{2111.14902}.

\bibitem{Caleb2022}
M.~{Caleb}, I.~{Heywood}, K.~{Rajwade}, M.~{Malenta}, B.~W. {Stappers},
  E.~{Barr}, W.~{Chen}, V.~{Morello}, S.~{Sanidas}, J.~{van den Eijnden},
  et~al., \href{http://dx.doi.org/10.1038/s41550-022-01688-x}{Nature Astronomy}
  \textbf{6}, 828 (2022), arXiv: \eprint{2206.01346}.

\bibitem{Nan2011}
R.~{Nan}, D.~{Li}, C.~{Jin}, Q.~{Wang}, L.~{Zhu}, W.~{Zhu}, H.~{Zhang},
  Y.~{Yue}, and L.~{Qian},
  \href{http://dx.doi.org/10.1142/S0218271811019335}{International Journal of
  Modern Physics D} \textbf{20}, 989 (2011), arXiv: \eprint{1105.3794}.

\bibitem{Li2018}
D.~{Li}, P.~{Wang}, L.~{Qian}, M.~{Krco}, P.~{Jiang}, Y.~{Yue}, C.~{Jin},
  Y.~{Zhu}, Z.~{Pan}, R.~{Nan}, et~al.,
  \href{http://dx.doi.org/10.1109/MMM.2018.2802178}{IEEE Microwave Magazine}
  \textbf{19}, 112 (2018), arXiv: \eprint{1802.03709}.

\bibitem{Brinkworth2004}
C.~S. {Brinkworth}, M.~R. {Burleigh}, G.~A. {Wynn}, and T.~R. {Marsh},
  \href{http://dx.doi.org/10.1111/j.1365-2966.2004.07538.x}{\mnras}
  \textbf{348}, L33 (2004), arXiv: \eprint{astro-ph/0312311}.

\bibitem{Harayama2013}
A.~{Harayama}, Y.~{Terada}, M.~{Ishida}, T.~{Hayashi}, A.~{Bamba}, and M.~S.
  {Tashiro}, \href{http://dx.doi.org/10.1093/pasj/65.4.73}{\pasj} \textbf{65},
  73 (2013).

\bibitem{Reding2020}
J.~S. {Reding}, J.~J. {Hermes}, Z.~{Vanderbosch}, E.~{Dennihy}, B.~C. {Kaiser},
  C.~B. {Mace}, B.~H. {Dunlap}, and J.~C. {Clemens},
  \href{http://dx.doi.org/10.3847/1538-4357/ab8239}{\apj} \textbf{894}, 19
  (2020), arXiv: \eprint{2003.10450}.

\bibitem{Pshirkov2020}
M.~S. {Pshirkov}, A.~V. {Dodin}, A.~A. {Belinski}, S.~G. {Zheltoukhov}, A.~A.
  {Fedoteva}, O.~V. {Voziakova}, S.~A. {Potanin}, S.~I. {Blinnikov}, and K.~A.
  {Postnov}, \href{http://dx.doi.org/10.1093/mnrasl/slaa149}{\mnras}
  \textbf{499}, L21 (2020), arXiv: \eprint{2007.06514}.

\bibitem{Caiazzo2021}
I.~{Caiazzo}, K.~B. {Burdge}, J.~{Fuller}, J.~{Heyl}, S.~R. {Kulkarni}, T.~A.
  {Prince}, H.~B. {Richer}, J.~{Schwab}, I.~{Andreoni}, E.~C. {Bellm}, et~al.,
  \href{http://dx.doi.org/10.1038/s41586-021-03615-y}{\nat} \textbf{595}, 39
  (2021), arXiv: \eprint{2107.08458}.

\bibitem{Buckley2017}
D.~A.~H. {Buckley}, P.~J. {Meintjes}, S.~B. {Potter}, T.~R. {Marsh}, and B.~T.
  {G{\"a}nsicke}, \href{http://dx.doi.org/10.1038/s41550-016-0029}{Nature
  Astronomy} \textbf{1}, 0029 (2017), arXiv: \eprint{1612.03185}.

\bibitem{Yao2017}
J.~M. {Yao}, R.~N. {Manchester}, and N.~{Wang},
  \href{http://dx.doi.org/10.3847/1538-4357/835/1/29}{\apj} \textbf{835}, 29
  (2017), arXiv: \eprint{1610.09448}.

\bibitem{Cordes2002}
J.~M. {Cordes} and T.~J.~W. {Lazio},
  \href{http://dx.doi.org/10.48550/arXiv.astro-ph/0207156}{arXiv e-prints}
  astro-ph/0207156 (2002), arXiv: \eprint{astro-ph/0207156}.

\bibitem{Ransom2002}
S.~M. {Ransom}, S.~S. {Eikenberry}, and J.~{Middleditch},
  \href{http://dx.doi.org/10.1086/342285}{\aj} \textbf{124}, 1788 (2002),
  arXiv: \eprint{astro-ph/0204349}.

\bibitem{Zhang2023}
L.~{Zhang}, P.~C.~C. {Freire}, A.~{Ridolfi}, Z.~{Pan}, J.~{Zhao}, C.~O.
  {Heinke}, J.~{Chen}, M.~{Cadelano}, C.~{Pallanca}, X.~{Hou}, et~al.,
  \href{http://dx.doi.org/10.3847/1538-4365/acfb03}{\apjs} \textbf{269}, 56
  (2023), arXiv: \eprint{2312.05835}.

\bibitem{Ng2015}
C.~{Ng}, D.~J. {Champion}, M.~{Bailes}, E.~D. {Barr}, S.~D. {Bates}, N.~D.~R.
  {Bhat}, M.~{Burgay}, S.~{Burke-Spolaor}, C.~M.~L. {Flynn}, A.~{Jameson},
  et~al., \href{http://dx.doi.org/10.1093/mnras/stv753}{\mnras} \textbf{450},
  2922 (2015), arXiv: \eprint{1504.08000}.

\bibitem{Lazarus2015}
P.~{Lazarus}, A.~{Brazier}, J.~W.~T. {Hessels}, C.~{Karako-Argaman}, V.~M.
  {Kaspi}, R.~{Lynch}, E.~{Madsen}, C.~{Patel}, S.~M. {Ransom}, P.~{Scholz},
  et~al., \href{http://dx.doi.org/10.1088/0004-637X/812/1/81}{\apj}
  \textbf{812}, 81 (2015), arXiv: \eprint{1504.02294}.

\bibitem{Singh2022}
S.~{Singh}, J.~{Roy}, U.~{Panda}, B.~{Bhattacharyya}, V.~{Morello}, B.~W.
  {Stappers}, P.~S. {Ray}, and M.~A. {McLaughlin},
  \href{http://dx.doi.org/10.3847/1538-4357/ac7b91}{\apj} \textbf{934}, 138
  (2022), arXiv: \eprint{2206.00427}.

\bibitem{Backer1970}
D.~C. {Backer}, \href{http://dx.doi.org/10.1038/228042a0}{\nat} \textbf{228},
  42 (1970).

\bibitem{Zhang2019}
L.~{Zhang}, D.~{Li}, G.~{Hobbs}, C.~H. {Agar}, R.~N. {Manchester},
  P.~{Weltevrede}, W.~A. {Coles}, P.~{Wang}, W.~{Zhu}, Z.~{Wen}, et~al.,
  \href{http://dx.doi.org/10.3847/1538-4357/ab1849}{\apj} \textbf{877}, 55
  (2019), arXiv: \eprint{1904.05482}.

\bibitem{Morello2020}
V.~{Morello}, E.~D. {Barr}, B.~W. {Stappers}, E.~F. {Keane}, and A.~G. {Lyne},
  \href{http://dx.doi.org/10.1093/mnras/staa2291}{\mnras} \textbf{497}, 4654
  (2020), arXiv: \eprint{2004.03701}.

\bibitem{Zhou2024}
D.~{Zhou}, P.~{Wang}, D.~{Li}, J.~{Fang}, C.~{Miao}, P.~C.~C. {Freire},
  L.~{Zhang}, D.~{Zhang}, H.~{Chen}, Y.~{Feng}, et~al.,
  \href{http://dx.doi.org/10.1007/s11433-023-2362-x}{Science China Physics,
  Mechanics, and Astronomy} \textbf{67}, 269512 (2024), arXiv:
  \eprint{2312.05868}.

\bibitem{Barsdell2012}
B.~R. {Barsdell}, M.~{Bailes}, D.~G. {Barnes}, and C.~J. {Fluke},
  \href{http://dx.doi.org/10.1111/j.1365-2966.2012.20622.x}{\mnras}
  \textbf{422}, 379 (2012), arXiv: \eprint{1201.5380}.

\bibitem{Lorimer2004}
D.~R. {Lorimer} and M.~{Kramer}, \emph{{Handbook of Pulsar Astronomy}},
  volume~4 (2004).

\bibitem{Jiang2020}
P.~{Jiang}, N.-Y. {Tang}, L.-G. {Hou}, M.-T. {Liu}, M.~{Kr{\v{c}}o}, L.~{Qian},
  J.-H. {Sun}, T.-C. {Ching}, B.~{Liu}, Y.~{Duan}, et~al.,
  \href{http://dx.doi.org/10.1088/1674-4527/20/5/64}{Research in Astronomy and
  Astrophysics} \textbf{20}, 064 (2020), arXiv: \eprint{2002.01786}.

\bibitem{Zhang2025}
L.~{Zhang}, F.~{Abbate}, D.~{Li}, A.~{Possenti}, M.~{Bailes}, A.~{Ridolfi},
  P.~C.~C. {Freire}, S.~M. {Ransom}, Y.-K. {Zhang}, M.~{Guo}, et~al.,
  \href{http://dx.doi.org/10.1016/j.scib.2025.03.022}{Science Bulletin}
  \textbf{70}, 1568 (2025), arXiv: \eprint{2503.08291}.

\bibitem{Gaia2016}
{Gaia Collaboration}, T.~{Prusti}, J.~H.~J. {de Bruijne}, A.~G.~A. {Brown},
  A.~{Vallenari}, C.~{Babusiaux}, C.~A.~L. {Bailer-Jones}, U.~{Bastian},
  M.~{Biermann}, D.~W. {Evans}, et~al.,
  \href{http://dx.doi.org/10.1051/0004-6361/201629272}{\aap} \textbf{595}, A1
  (2016), arXiv: \eprint{1609.04153}.

\bibitem{Gaia2023}
{Gaia Collaboration}, A.~{Vallenari}, A.~G.~A. {Brown}, T.~{Prusti}, J.~H.~J.
  {de Bruijne}, F.~{Arenou}, C.~{Babusiaux}, M.~{Biermann}, O.~L. {Creevey},
  C.~{Ducourant}, et~al.,
  \href{http://dx.doi.org/10.1051/0004-6361/202243940}{\aap} \textbf{674}, A1
  (2023), arXiv: \eprint{2208.00211}.

\bibitem{Hotan2004}
A.~W. {Hotan}, W.~{van Straten}, and R.~N. {Manchester},
  \href{http://dx.doi.org/10.1071/AS04022}{\pasa} \textbf{21}, 302 (2004),
  arXiv: \eprint{astro-ph/0404549}.

\bibitem{Sault1995}
R.~J. {Sault}, P.~J. {Teuben}, and M.~C.~H. {Wright}, in \emph{Astronomical
  Data Analysis Software and Systems IV}, (edited by R.~A. {Shaw}, H.~E.
  {Payne}, and J.~J.~E. {Hayes}), volume~77 of \emph{Astronomical Society of
  the Pacific Conference Series}, 433 (1995), arXiv: \eprint{astro-ph/0612759}.

\bibitem{Offringa2014}
A.~R. Offringa, B.~McKinley, Hurley-Walker, et~al.,
  \href{http://dx.doi.org/10.1093/mnras/stu1368}{MNRAS} \textbf{444}, 606
  (2014).

\bibitem{Pritchard2024b}
J.~Pritchard, askap-vast/dstools: v1.0.0 (2024),
  \urlprefix\url{https://doi.org/10.5281/zenodo.13626183}.

\bibitem{Marcote2017}
B.~{Marcote}, T.~R. {Marsh}, E.~R. {Stanway}, Z.~{Paragi}, and J.~M.
  {Blanchard}, \href{http://dx.doi.org/10.1051/0004-6361/201730948}{\aap}
  \textbf{601}, L7 (2017), arXiv: \eprint{1705.00600}.

\bibitem{cumming02}
A.~{Cumming}, \href{http://dx.doi.org/10.1046/j.1365-8711.2002.05434.x}{\mnras}
  \textbf{333}, 589 (2002), arXiv: \eprint{astro-ph/0202079}.

\bibitem{zhang00}
B.~{Zhang}, A.~K. {Harding}, and A.~G. {Muslimov},
  \href{http://dx.doi.org/10.1086/312542}{\apjl} \textbf{531}, L135 (2000),
  arXiv: \eprint{astro-ph/0001341}.

\bibitem{ferrario20}
L.~{Ferrario}, D.~{Wickramasinghe}, and A.~{Kawka},
  \href{http://dx.doi.org/10.1016/j.asr.2019.11.012}{Advances in Space
  Research} \textbf{66}, 1025 (2020), arXiv: \eprint{2001.10147}.

\bibitem{rea24}
N.~{Rea}, N.~{Hurley-Walker}, C.~{Pardo-Araujo}, M.~{Ronchi}, V.~{Graber},
  F.~{Coti Zelati}, D.~{de Martino}, A.~{Bahramian}, S.~J. {McSweeney}, T.~J.
  {Galvin}, et~al., \href{http://dx.doi.org/10.3847/1538-4357/ad165d}{\apj}
  \textbf{961}, 214 (2024), arXiv: \eprint{2307.10351}.

\bibitem{reisenegger09}
A.~{Reisenegger}, \href{http://dx.doi.org/10.1051/0004-6361/200810895}{\aap}
  \textbf{499}, 557 (2009), arXiv: \eprint{0809.0361}.

\bibitem{tout08}
C.~A. {Tout}, D.~T. {Wickramasinghe}, J.~{Liebert}, L.~{Ferrario}, and J.~E.
  {Pringle}, \href{http://dx.doi.org/10.1111/j.1365-2966.2008.13291.x}{\mnras}
  \textbf{387}, 897 (2008), arXiv: \eprint{0805.0115}.

\bibitem{isern17}
J.~{Isern}, E.~{Garc{\'\i}a-Berro}, B.~{K{\"u}lebi}, and
  P.~{Lor{\'e}n-Aguilar},
  \href{http://dx.doi.org/10.3847/2041-8213/aa5eae}{\apjl} \textbf{836}, L28
  (2017), arXiv: \eprint{1702.01813}.

\bibitem{beniamini23}
P.~{Beniamini}, Z.~{Wadiasingh}, J.~{Hare}, K.~M. {Rajwade}, G.~{Younes}, and
  A.~J. {van der Horst}, \href{http://dx.doi.org/10.1093/mnras/stad208}{\mnras}
  \textbf{520}, 1872 (2023), arXiv: \eprint{2210.09323}.

\bibitem{wadiasingh19}
Z.~{Wadiasingh} and A.~{Timokhin},
  \href{http://dx.doi.org/10.3847/1538-4357/ab2240}{\apj} \textbf{879}, 4
  (2019), arXiv: \eprint{1904.12036}.

\bibitem{suvorov23}
A.~G. {Suvorov} and A.~{Melatos},
  \href{http://dx.doi.org/10.1093/mnras/stad274}{\mnras} \textbf{520}, 1590
  (2023), arXiv: \eprint{2301.08541}.

\bibitem{thompson01}
C.~{Thompson} and R.~C. {Duncan}, \href{http://dx.doi.org/10.1086/323256}{\apj}
  \textbf{561}, 980 (2001), arXiv: \eprint{astro-ph/0110675}.

\bibitem{akgün16}
T.~{Akg{\"u}n}, J.~A. {Miralles}, J.~A. {Pons}, and P.~{Cerd{\'a}-Dur{\'a}n},
  \href{http://dx.doi.org/10.1093/mnras/stw1762}{\mnras} \textbf{462}, 1894
  (2016), arXiv: \eprint{1605.02253}.

\bibitem{Philippov2018}
A.~A. {Philippov} and A.~{Spitkovsky},
  \href{http://dx.doi.org/10.3847/1538-4357/aaabbc}{\apj} \textbf{855}, 94
  (2018), arXiv: \eprint{1707.04323}.

\bibitem{Takata2017}
J.~{Takata}, H.~{Yang}, and K.~S. {Cheng},
  \href{http://dx.doi.org/10.3847/1538-4357/aa9b33}{\apj} \textbf{851}, 143
  (2017), arXiv: \eprint{1712.03488}.

\bibitem{Tauris1998}
T.~M. {Tauris} and R.~N. {Manchester},
  \href{http://dx.doi.org/10.1046/j.1365-8711.1998.01369.x}{\mnras}
  \textbf{298}, 625 (1998).

\bibitem{Kolonko2004}
M.~{Kolonko}, J.~{Gil}, and K.~{Maciesiak},
  \href{http://dx.doi.org/10.1051/0004-6361:20034399}{\aap} \textbf{428}, 943
  (2004), arXiv: \eprint{astro-ph/0412159}.

\end{thebibliography}

\end{multicols}
\end{document}